\documentclass[aps,twocolumn,showpacs]{revtex4}
\usepackage{graphicx}
\begin{document}
\title{Dynamics of   Crossover from a Chaotic to a Power Law State in
Jerky Flow}
\author{M.S. Bharathi$^{1}$ and G. Ananthakrishna$^{1,2}$}
\affiliation{$^1$ Materials Research Centre, Indian Institute of Science,
Bangalore-560012, India\\
$^{2}$ Centre for Condensed Matter Theory, Indian Institute of Science,
Bangalore-560012, India}

\begin{abstract}

We study the dynamics of an intriguing crossover from a chaotic to
a power law state as a function of strain rate within the context
of a recently introduced model which reproduces the crossover.
While the chaotic regime has a small set of positive Lyapunov
exponents, interestingly, the scaling regime has a power law
distribution of null exponents which also exhibits a power law.
The slow manifold analysis of the model shows that while a large
proportion of dislocations are pinned in the chaotic regime, most
of them are pushed to the threshold of unpinning in the scaling
regime, thus providing insight into the mechanism of crossover.

\end{abstract}
\maketitle

The Portevin Le-Chatelier  effect \cite{PLC}, or the jerky flow is
a rare example where the complex spatio-temporal dynamics results
from the collective behavior of participating defects, namely,
dislocations \cite{Set}. When samples (usually thin strips) of
dilute alloys are subjected constant strain rate deformation in a
window of applied strain rates ($\dot\epsilon_a$) and
temperatures, one finds repeated stress drops (or yield drops).
Each stress drop is associated with the formation of a band of
dislocations. At low strain rates the bands are static nucleating
randomly in space. At intermediate $\dot\epsilon_a$, bands
nucleate one ahead of the other in a hopping manner. At high
strain rates, bands propagate continuously. The bands found in
different regimes of strain rates  are considered to be different
correlated states of dislocations within the bands \cite{Bhar01}.

A classical explanation of the jerky flow is through dynamic
strain aging \cite{Set}.  At small velocities of dislocations (or
$\dot\epsilon_a$), solute atoms have sufficient time to diffuse to
dislocations, and  pin them. Longer they are arrested, larger is
the stress required to unpin  them. As the stress increases,
dislocations get unpinned  and move fast till they are again
pinned due to diffusing solutes and other pinning centers. The
process repeats itself.  Further, the competition between the time
scales associated with diffusion and dislocation mobility
translates, at the macroscopic level, to a negative strain rate
sensitivity of the flow stress. This intermittent sequence of
loading and unloading, and the negative flow rate sensitivity are
typical features of many stick-slip situations such as fault
dynamics \cite{Burridge}, frictional sliding \cite{Persson} and
peeling of an adhesive tape \cite{Maug} and charge density waves
\cite{Tang87,Dumas}. ( Indeed, the similarity of the PLC effect
with charge density waves has been studied in detain in Ref.
\cite{Dumas}.) The power law statistics of stress drops
\cite{Bhar01,Anan99} at {\it high} $\dot{\epsilon}_a$ is similar
to those in earthquakes \cite{Burridge} and many other power law
systems \cite{Bak,Jensen} which however are seen at {\it low}
drives.

This rich spatio-temporal dynamics  has defied a proper
understanding due to lack of techniques to describe the collective
behavior of dislocations \cite{Set}. Recent studies  using
nonlinear dynamical methods have  shown that a rich body of
dynamical correlations is hidden in the stress-strain curves of
jerky flow \cite{Noro97}. More recently, {\it an intriguing
crossover} from a chaotic  to power law regime has been observed
as a function of strain rate \cite{Anan99,Bhar01}. As the
crossover is seen in single and poly-crystals, it appears to be
insensitive to the microstructure.

This crossover is unusual in a number of ways. First, it is a {\it
rare example of a transition between two dynamically distinct
states.} Chaos is characterized by self similarity of the
attractor and sensitivity to  initial conditions arising from a
few degrees of freedom. In contrast, power laws are scale free and
are infinite dimensional. Thus, most systems exhibit {\it either
of these regimes}. To the best of our knowledge, this is one of
the two experimental situations where both are seen in the same
system, the other being in hydrodynamic turbulence \cite{Lib}.
Second, the power law in the PLC effect ( as also in turbulence)
is seen at {\it high drive rates}.  On the other hand, power laws
observed in varied systems, usually seen in {\it slowly driven}
dissipative systems, are conventionally explained by invoking the
concept of self-organized criticality (SOC)
\cite{Bak,Bak96,Jensen}. Thus, one suspects that the dynamical
features of the power law here to be closer to turbulence than the
conventional SOC systems as we will show.

Recently, we have succeeded in explaining this crossover in the
dynamics of the PLC effect by extending  the dynamical model for
the PLC effect due to Ananthakrishna and coworkers
\cite{Anan82,Bhar02}. The extended model also explains the three
types of dislocation bands seen with increasing strain rate
\cite{Bhar03}. However, the dynamics of the crossover has not been
elucidated. A natural tool for characterizing the crossover is to
follow the Lyapunov spectrum as a function of strain rate. This
will be supplemented by  the slow-manifold approach
\cite{Rajesh,Rajesh00} which allows us get a geometrical picture
of the changes occurring in the configuration of dislocations
during crossover.

We shall briefly describe the extended dynamical \cite{Bhar02}
model. The original dynamical model \cite{Anan82} captures the
well separated time scales implied in dynamic strain aging by
using the fast mobile $\rho_m(x,t)$, immobile $\rho_{im}(x,t)$,
and Cottrell type of dislocations $\rho_c(x,t)$. Then, all
qualitative features of the effect emerge due to nonlinear
interaction of  these populations, assumed to represent collective
degrees of freedom. In spite of the idealized nature of the model,
it has been successful in explaining several generic features of
the effect, notably - the occurrence of the effect in a window  of
strain rates ( or temperatures), and the emergence of negative
strain rate sensitivity of the flow stress \cite{Anan82,Rajesh}.
The model also predicts {\it chaotic stress drops} \cite{Anan83}
which has been subsequently verified by analyzing experimental
signals \cite{Noro97,Anan99}, and hence has the right framework to
model the crossover.  The equation of motion of the scaled
dislocation densities \cite{Anan82,Rajesh}  $\rho_m(x,t)$,
$\rho_{im}(x,t)$ and $\rho_c(x,t)$ are given by :
\begin{eqnarray}
\nonumber
\dot{\rho}_{m}& =&  -b_0\rho_m^2 -\rho_m\rho_{im} +\rho_{im} - a \rho_m +
\phi_{eff}^m\rho_m \\
&+&  ({{\cal{D}}/\rho_{im}})(\phi_{eff}^m(x)\rho_m)_{xx},\\
\dot{\rho}_{im} & = & b_0(b_0\rho_m^2
-\rho_m\rho_{im}  -\rho_{im}+a\rho_c), \\
\dot{\rho}_c & = & c(\rho_m-\rho_c).
\end{eqnarray}
The overdot and the subscript $x$ refer respectively to the time
and space derivatives. The first term in Eq.(1), $b_0\rho_m^2$,
refers to the immobilization of two mobile dislocations due to
formation of locks, $\rho_m\rho_{im}$  to the annihilation of a
mobile dislocation with an immobile one,    $\rho_{im}$ to the
remobilization of the immobile dislocation due to stress or
thermal activation, and  $a\rho_m$  represents the immobilization
of mobile dislocation due to the aggregation of solute atoms. Once
a mobile dislocation starts acquiring  solute atoms,  we regard it
as the Cottrell type $ \rho_c$. They eventually become immobile as
more solute atoms aggregate. The aggregation  of solute atoms can
be regarded as the definition of $\rho_c$, ie., $\rho_c =
\int_{-\infty}^t dt^{\prime}\rho_m(t^{\prime} ) exp [- c(t
-t^{\prime})]$. (See Ref. \cite{Rajesh}.) $\phi_{eff}^m\rho_m $
refers to the rate of production of dislocations due to cross
glide. This depends on the velocity of mobile dislocations taken
to be $ V_m (\phi) = \phi_{eff}^m$, where $\phi_{eff} = (\phi - h
\sqrt{\rho_{im}})$ is the scaled effective stress, $\phi$ the
scaled stress, $m$ the velocity exponent and $h$ a work hardening
parameter. In the original model cross-slip has been used as a
source of dislocation multiplication which, however, is
intrinsically nonlocal. During cross-slip  dislocations leave the
slip plane due to, for instance, the effect of repulsive internal
stresses and then slip back onto the slip plane. This mechanism
transports dislocation densities through space. This is known to
lead to a 'diffusive' type coupling \cite{Set}. Let $\Delta x$ be
an elementary length. Then, the flux $\Phi(x)$ flowing from $x \pm
\Delta x$ and out of $x$ is given by $\Phi(x) + \frac{p}{2}
\left[\ \Phi(x+\Delta x) - 2 \Phi(x) + \Phi(x - \Delta x)\right]$,
where $\Phi(x) = \rho_m(x)\phi^{m}_{eff}(x)$. Here $p$ is the
probability of spreading into neighboring elements. Expanding
$\Phi(x \pm \Delta x)$ up to the leading terms, we get $\rho_m
\phi^{m}_{eff} + \frac{p}{2}\frac {\partial^2(\rho_m
\phi^{m}_{eff})}{\partial x^2} (\Delta x)^2.$ Since cross-slip
spreads only into regions of minimum back stress arising from $\rho_{im}$
ahead of it, we use $\Delta x^2 = <\Delta
x^2> = \bar {r}^2 \rho_{im}^{-1}$. Here $<.>$ refers to the
ensemble average and $\bar{r}$ is an elementary (dimensionless)
length with ${\cal{D}}=p \bar r^2/2$. The scaled constants, $a$, $b_0$ and
$c$ refer, respectively, to the concentration of solute atoms
slowing down the mobile dislocations, the reactivation of immobile
dislocations, and the diffusion  rate of solute atoms. The orders
of magnitudes of these constants are known from the basic
mechanisms and their correspondence with experimental quantities
\cite{Rajesh}. Defining  $\dot\epsilon$, $d$ and $l$ as  the scaled strain rate,
effective modulus of the machine and the sample, and
  length of the sample, respectively, the machine equation reads
\begin{equation}
\dot{\phi}= d [\dot{\epsilon}-({1}/{l})\int_0^l\rho_m(x,t).
\phi_{eff}^m(x,t)dx],
\end{equation}
Global coupling in Eq. (4) is similar to studies on space charge currents in
semiconductors where  the integrated electric field balances the
applied voltage.

In the domain of the crossover parameter, $\dot \epsilon$, our
earlier analysis \cite{Rajesh,Rajesh00} shows that the effect is
observed between a forward Hopf bifurcation at low strain rate and
a reverse one at high  $\dot \epsilon$. The reverse Hopf
bifurcation {\it implies decreasing amplitudes of stress drops as
in experiments}. All the interesting dynamics, including chaos, is
seen in this regime as shown in Ref. \cite{Rajesh,Rajesh00}.

We discretize the  specimen length into $N$ equal parts, and solve
for $\rho_m(j,t)$, $\rho_{im}(j,t), \rho_c(j,t)$, $j= 1,...N$, and
${\phi}(t)$. Due to the widely differing time scales, appropriate
care is taken in the numerical solutions by using a variable step
fourth order Runge-Kutta scheme with an accuracy of $10^{-6}$ for
all the variables. The spatial derivative  is approximated by its
central difference. The initial values are taken as the steady
state values for the variables (as the long term evolution does
not depend on the initial values) with a Gaussian spread along the
length of the sample. In experiments, the ends of the sample have
large strains induced due to  high stress concentrations at the
grips. To mimic the strain, we set $\rho_{im}(j,t)$, $j=1$ and $N$
to values two orders higher than the  rest of the sample.
Further, as bands cannot propagate into the grips, we use
$\rho_m(j,t) =\rho_c(j,t)=0$ at $j =1$  and $N$. For the numerical
work, we use $a = 0.8, b_0 = 0.0005, c = 0.08, d =0.00006, m =
3.0, h=0$ and ${\cal{D}}=0.5$. However, the results hold true for a wide
range of parameters values in the instability domain including
that of ${\cal{D}}$. For these values, the PLC effect is seen in the range
$ 10 < \dot \epsilon < 2000$. Chaotic stress drops are seen at low
strain rates and power law statistics at high $\dot \epsilon$
\cite{Bhar02}.

We identify the chaotic regime by calculating the Lyapunov
exponents,  $\lambda_i$ $(i=1,..,M=3N+1)$,  using Eqs. (1-4). (The
various systems sizes studied from $N=100-3333$ show a rapid
convergence of the results even around 300.)  The largest Lyapunov
exponent, $\lambda_1$, is obtained by averaging over 15000 time
steps after stabilization. $\lambda_1$ becomes positive at
$\dot\epsilon = 35$, reaching a maximum at $\dot \epsilon=120$,
and practically vanishes around 250. (See Fig. 3a of Ref.
\cite{Bhar02}. Periodic states are seen prior to chaos.) In the
chaotic region, the distribution of Lyapunov exponents is quite
broad. A plot for $\dot \epsilon =120$ is shown Fig. 1a. Of these,
only  6.2\% of $M(=1051)$ are positive. As $\dot\epsilon$
increases to 280, concomitant with the decrease in the maximum
Lyapunov exponent to a small value, $\approx 5.2 \times 10^{-4}$,
the number of null exponents (almost vanishing) {\it increases
gradually} reaching a value $\approx 0.38M$ in the range
$[-0.00052, 0.00052]$ (compared to only a few for $\dot\epsilon
=120$).  For $\dot\epsilon \ge 250$, below a resolution $\sim
10^{-4}$, even as the first few exponents are distinguishable,
most cross each other as a function of time, but the (time
averaged) distribution remains unaffected. The finite density of
null exponents has a peaked nature in the interval $250 \le
\dot\epsilon \le 700$. A  plot is shown in Fig. 1c  for $\dot
\epsilon=280$ which can be fitted to a power law $D(\vert \lambda
\vert) \sim \vert \lambda \vert^{- \gamma}$ as shown in Fig. 1c
with $\gamma = 0.6$. It may be pertinent to state here that the
uncorrelated bands and hopping bands are seen in the chaotic
regime while the continuously propagating bands are seen at high
strain rate \cite{Bhar03}.

\begin{figure}[!t]
\includegraphics[height=3.2cm,width=8.5cm]{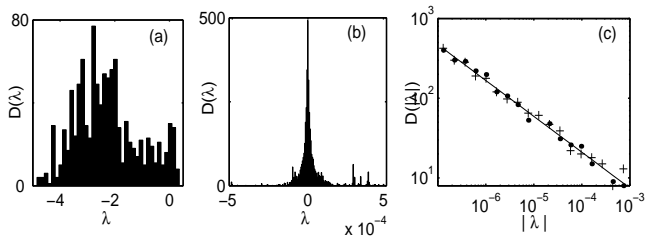}
\caption{(a) Distribution of the Lyapunov exponents at
$\dot\epsilon=120$ for $M=1051$. (b) and (c) Distribution of the
null exponents at $\dot\epsilon=280$ for $M=10000$. In (c) (+)
refers to positive and ($\bullet$) to negative null Lyapunov
exponents.}
\end{figure}

As yield drops are caused by unpinning of dislocations from pinned
configuration, we first need to identify these configurations.
This can be done by using the slow manifold approach. Here we
recall some relevant results \cite{Rajesh,Rajesh00} on the slow
manifold  of the original model ($D=0$) for further use when $D
\neq 0$. Slow manifold expresses the fast variable in terms of the
slow variables, conventionally done  by setting the derivative of
the fast variable to zero \cite{Rajesh,Rajesh00}. Here, $\dot
\rho_m = 0$ gives $\rho_m =\rho_m(\rho_{im},\phi)$. Instead, we
use $\rho_m$ in terms of a single slow variable $\delta  = \phi^m
- \rho_{im} -a$. We note that $\delta$ takes on small positive and
negative values as both $\rho_{im} $ and $\phi$ are small and
positive. Using $\dot \rho_m = g(\rho_m,\phi) = - b_0 \rho_m^2 +
\rho_m \delta +\rho_{im} =0$, and noting that $\rho_m > 0$, we get
two solutions $\rho_m = [\delta + (\delta^2 + 4 b_0
\rho_{im})^{1/2}]/2b_0$, one for $\delta <0 $ and another $\delta
> 0$. For regions of $\delta < 0$, as $b_0$ is small  $\sim
10^{-4}$, we get $\rho_m/\rho_{im} \approx - 1/\delta$ which takes
on small values defining a part of the slow manifold, $S_2$. Since
physically pinned configuration of dislocations implies small
mobile density and large immobile density, we refer to the region
of $S_2$ as the 'pinned state of dislocations'. Further, {\it
larger negative values of  $\delta$ correspond to strongly pinned
configurations,} as they refer to smaller ratio of
$\rho_m/\rho_{im}$. Corresponding to $\delta > 0$, another
connected piece $S_1$ is defined by {\it large values} of
$\rho_m$, given by $\rho_m \approx  \delta/b_0$, which we refer to
as the 'unpinned state of dislocations'.  $S_2$ and $S_1$ are
separated by $\delta = 0$, which we refer to as {\it the fold
line} \cite{Rajesh,Rajesh00}(see below).
\begin{figure}[!t]
\mbox{
\includegraphics[height=3.5cm,width=5.0cm]{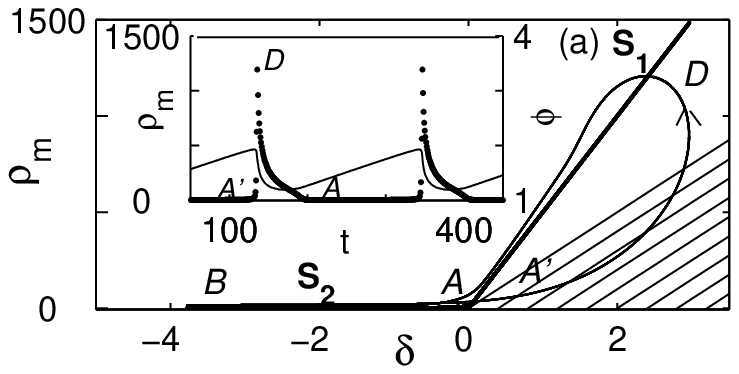}
\includegraphics[height=3.5cm,width=3.5cm]{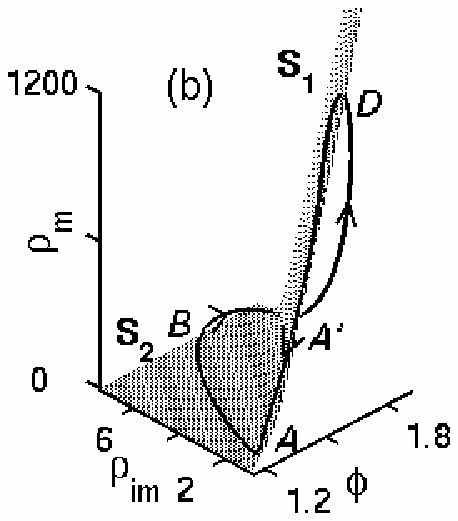}}
\caption{(a) Bent slow manifold $S_1$ and $S_2$ (thick lines) with
a simple trajectory for $\dot\epsilon=200$ and $m= 3$. Inset:
$\rho_m$ (dotted curve) and $\phi$. (b) Same trajectory in the
$(\phi,\rho_{im},\rho_m)$ space.}
\end{figure}
\noindent A plot of the slow manifold in the $\delta-\rho_m$ plane
is shown in Fig. 2a along with a simple monoperiodic trajectory
describing the changes in the densities during one
loading-unloading cycle.   The inset shows $\rho_m(t)$ and
$\phi(t)$. For completeness, the corresponding plot of the slow
manifold in the $(\rho_m,\rho_{im},\phi)$ space is shown in Fig.
2b, along with the trajectory and the symbols. Note that $S_2$ and
$S_1$ are separated by $\delta =\phi^m-\rho_{im} -a = 0$, and
hence the name {\it fold line}.  In Fig. 2a, as the trajectory
enters $S_2$ at $A$ and moves into $S_2$, $\delta$ takes  a
maximum negative value at $B$. Then $\delta$ increases as the
trajectory returns to $A^{\prime}$ before leaving $S_2$. The
corresponding segment is $ABA^{\prime}$ in Fig. 2b, which is
identified with the  flat region of $\rho_m(t)$ in the inset of
Fig. 2a. As the trajectory crosses $\delta =0$, $\partial
g/\partial \rho_m $ becomes positive and it accelerates into the
shaded region (Fig. 2a) rapidly till  it reaches $\rho_m =
\delta/2b_0$. Thereafter it settles down quickly on $S_1$
decreasing rapidly till it reenters $S_2$ again at $A$. The burst
in $\rho_m$ (inset in Fig. 2a) corresponds to the segment
$A^{\prime}DA$ in Fig. 2a and b.

Now consider the stress changes as the state of  the system goes
though a burst of plastic activity.  For ${\cal{D}} =0$, Eq.(4) reduces to
$\dot \phi = d[ \dot \epsilon -\dot {\epsilon}_p]$, where
$\dot\epsilon_p = \phi^m \rho_m$ defines the plastic strain rate.
Since $\rho_m$ is small  and nearly constant on $S_2$, stress
increases monotonically. However, during the burst in $\rho_m$
($A^{\prime} D A$ in the inset), $\dot \epsilon_p (t)$ exceeds
$\dot \epsilon$ leading to an yield drop. Since $\rho_m$ grows
outside $S_2$, $\delta =0$ separates the pinned state from the
unpinned state, and hence $\delta =0$ physically corresponds to
the value of the effective stress at which dislocations are
unpinned.

Now, we extend the slow manifold analysis to the case when spatial
coupling $D \neq 0$ to study the changes in the spatial configuration of 
dislocations as we go from the chaotic to scaling regime. Using 
$h=0, \phi_{eff}=\phi$,  the plastic strain rate  $\dot \epsilon_p(t)$ is given by
$\dot \epsilon_p(t)
= \phi^m(t)\frac{1}{l}\int_0^l\rho_m(x,t)dx = \phi^m(t) \bar\rho_m(t)$,
where $\bar \rho_m(t)$ is the mean mobile
density $(=\sum_j\rho_m(j,t)/N)$. Thus,  the yield drop is
controlled by the spatial average $\bar \rho_m(t)$ and not by individual values of
$\rho_m(j)$. Since the yield drop occurs when $\bar \rho_m(t)$  grows
rapidly, it is adequate to examine the spatial
configurations on the slow manifold  at the onset and at the end of
typical yield drops.
\begin{figure}[!t]
\includegraphics[height=5.7cm,width=8.5cm]{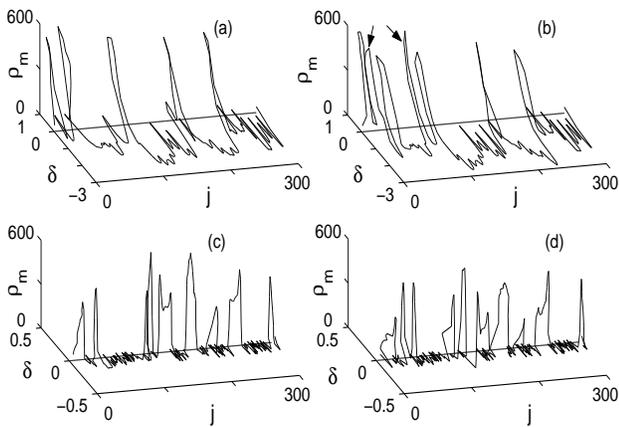}
\caption{Dislocation configurations on the slow manifold at the
inset and at the end of an yield drop: (a) and (b) for
$\dot\epsilon=120$ (chaos), and (c) and (d) for $\dot\epsilon=280$
(scaling).}
\end{figure}
Figures 3 a,b and 3 c,d  show respectively, plots of $j,\delta
(j),\rho_m(j)$ for the chaotic state $\dot \epsilon = 120$  and
the power law state $\dot \epsilon =280$, at the onset and at the
end of an yield drop. It is clear that for $\dot \epsilon=120$,
both at the onset and at the end of a typical large yield drop
(Fig. 3a,b), most $\rho_m(j)$s are small with large negative
values of $\delta(j)$, i.e., most dislocations are in {\it a
strongly pinned state}. The arrows show the increase in
$\rho_m(j)$ at the end of the yield drop. In contrast, in the
scaling regime, for $\dot \epsilon = 280$, {\it most dislocations
are at the threshold of unpinning with $\delta(j) \approx 0$,}
both at the onset and end of the yield drop (Fig. 3c,d). This also
implies that they remain close to this threshold {\it all the
time} (Fig. 3d).
Since $\delta(j) \approx 0$ (for most $j$'s) refers to a
marginally stable state, it can produce almost any response. This
in turn implies that the magnitudes of yield drops $\Delta \phi$
are scale free. We have verified that the edge-of-unpinning
picture is valid in the entire scaling regime for a range of $N =
100 - 1000$. Further, the number of spatial elements reaching the
threshold of unpinning $\delta = 0$, during an yield drop
increases as we approach the scaling regime.

Several comments may be in order on the dynamics of the crossover.
First, the crossover itself is smooth as the changes in the
Lyapunov spectrum are gradual. Second, the power law here is of
purely dynamical origin. This is a result of the reverse Hopf
bifurcation at high strain rates
 which limits the average  stress drop
amplitude to small values \cite{Rajesh,Rajesh00}. Third, our
analysis shows that the power law regime of stress drops occurring
at high strain rates belongs to a different universality class
as it is characterized by a dense set of null exponents. As zero
exponents correspond to a marginal situation, their finite density
physically implies that most spatial elements are close to
criticality. This is supported by the {\it geometrical picture of
the slow manifold} where most dislocations are at the threshold of
unpinning, $\delta=0$. In contrast, going by the few reports, the
marginal nature of conventional SOC models  does not display any
characteristic feature  in the Lyapunov spectrum  \cite{Erzan}.
(For instance, no zero and positive exponents, a positive and zero
exponent, zero exponent in the large $N$ limit etc have been
reported \cite{Erzan}.) More significantly, the dense set of null
Lyapunov exponents themselves follow a power law. Further, we note
that the Lyapunov spectrum evolves from a set of both positive and
negative, but few null exponents in the chaotic region, to a dense
set of marginal exponents as we reach the power law regime. Thus,
the dense set of null exponents in our model is actually similar
to that obtained in shell models of turbulence where the power law
is seen at high drive values \cite{Yamada}. However, there are
significant differences. First, we note that the shell model
\cite{Yamada} cannot explain the crossover as it is only designed
to explain the power law regime. Further, the maximum Lyapunov
exponent is large for small viscosity parameter $(\lambda_1
\propto viscosity^{-1/2})$ in shell models \cite{Yamada} in
contrast to near zero value in our model.

In conclusion we have demonstrated that the changes in the
Lyapunov spectrum provides a good insight into the dynamical
mechanism controlling the crossover. The slow manifold analysis,
applied for the first time to study the crossover, is particularly
useful in giving a geometrical picture of the spatial
configurations in the chaotic and scaling regimes. This picture
explains the origin of small amplitude stress drops at high strain
rates.
 Finally, as far as we know, this is first fully dynamical
model which exhibits a crossover from chaotic to power law regime
and should be of interest to the area of dynamical systems.

This work is supported by Department of Science and Technology,
New Delhi, India.


\begin{thebibliography}{99}
\bibitem{PLC} F. Le Chatelier,  Rev. de M\'etallurgie {\bf 6}, 914 (1909).
\bibitem{Set}L.P. Kubin, C. Fressengeas and G. Ananthakrishna, in {\it Collective Behavior of
Dislocations}, edited by F.R.N. Nabarro and M.S. Deusbery, {\it Dislocations in
Solids} Vol.11 (North-Holland, Amsterdam, 2002).
\bibitem{Bhar01}M.S. Bharathi, {\it et al.}, Phys.Rev.Lett. {\bf 87},
165508 (2001).
\bibitem{Burridge}R. Burridge and L. Knopoff, Bull. Seismol. Soc. Am.
{\bf 57}, 3411 (1967).
\bibitem{Persson} B.N.J. Persson and E. Tosatti, {\it Physics of Sliding
Friction}, (Kluwer Academic Publishers, Dordrecht, 1996).
\bibitem{Maug}D. Maugis and M. Barquins, {\it Adhesion}, edited by K.W.
Allen, Vol. 12. (Elsevier, London, 1988).
\bibitem{Tang87} C. Tang {\it et. al.,} Phys. Rev. Lett. {\bf 58},
1161 (1987).
\bibitem{Dumas}J. Dumas and D. Feinberg, Europhys. Lett. {\bf 2},
555 (1986).
\bibitem{Anan99}G. Ananthakrishna  {\it et al.}, Phys. Rev. E. {\bf 60}
5455 (1999).
\bibitem{Bak}P. Bak, C. Tang and K. Wiesenfeld,  Phys. Rev. Lett. {\bf
59}, 381 (1987); Phys. Rev. A. {\bf 38}, 364 (1988).
\bibitem{Jensen}H.J. Jensen, {\it Self-Organized Criticality}. (Cambridge
University Press, Cambridge, 1998).
\bibitem{Noro97} S.J. Noronha, {\it et al.}, Int. J. of Bifurcation and
Chaos {\bf 7} 2577 (1997) and the references therein.
\bibitem{Lib} See for instance F. Heslot,  B. Castaing and A. Libchaber, Phys. Rev. A. {\bf 36}, 5780 (1987).
\bibitem{Bak96}P. Bak, {\it How Nature Works}, (Springer - Verlag,
NewYork, 1996).
\bibitem{Anan82}G. Ananthakrishna and M.C. Valsakumar, J. Phys. D. {\bf
15}, L171 (1982).
\bibitem{Bhar02}M.S. Bharathi and G. Ananthakrishna, Europhys. Lett. {\bf 60}, 234 (2002).
\bibitem{Bhar03}M.S. Bharathi, S. Rajesh and G. Ananthakrishna, Scripta Mater. {\bf 48}, 1355 (2003).
\bibitem{Rajesh} S. Rajesh and G. Ananthakrishna, Phys. Rev. E. {\bf 61},
3664 (2000).
\bibitem{Rajesh00}S. Rajesh and G.Ananthakrishna, Physica D {\bf 140}, 193
(2000).
\bibitem{Anan83}G. Ananthakrishna and M.C. Valsakumar, Phys. Lett. {\bf
A95}, 69 (1983).

\bibitem{Erzan} A. Erzan and S. Sinha, Phys. Rev. Lett. {\bf 66}, 2750
(1991); M. de Sousa Vieira and A.J. Lichtenberg, Phys. Rev. E.
{\bf 53}, 1441 (1996); B. Cessac, Ph. Blanchard and T. Kruger,
Phys. Rev. E. {\bf 64}, 016133 (2001).
\bibitem{Yamada} M. Yamada and K. Ohkitani, J. Phys. Soc. Jpn. {\bf 56}, 4210
(1987); A. Cristani {\it et al}, Physica D, {\bf 76}, 239 (1994).

\end{thebibliography}
\end{document}